# OSDG – Open-Source Approach to Classify Text Data by UN Sustainable Development Goals (SDGs)

**Authors:** Lukas Pukelis[*,1], Núria Bautista Puig[2], Mykola Skrynik[1], Vilius Stanciauskas[1]

## Abstract

*Sustainable Development Goals (SDGs) bring together the diverse development community and provide a clear set of development targets for 2030. Given a large number of actors and initiatives related to these goals, there is a need to have a way to accurately and reliably assign text to different input: scientific research, research projects, technological output or documents to specific SDGs. In this paper we present Open Source SDG (OSDG) project and tool which does so by integrating existing research and previous classification into a robust and coherent framework. This integration is based on linking the features from the variety of previous approaches, like ontology items, keywords or features from machine-learning models, to the topics in [Microsoft Academic Graph](Microsoft Academic Graph).*

[1] Public Policy and Management Institute (PPMI), Vilnius, Lithuania
[2] Laboratory for Metric Information Studies (LEMI), University Carlos III of Madrid (UC3M)
* lukas@technote.ai

## Introduction

Sustainable Development Goals (SDGs) were adopted by the UN in 2015 as a set of reference goals for the Sustainable Development Community for the period 2015-2030. These goals serve two main purposes: first, they serve as the focal points highlighting the most salient issues in the more general theme of sustainable development. Second, they indicate a set of targets to which the global community benchmarks its progress over this period (UN 2020). Different from its predecessor, the Millennium Development Goals, SDGs require involvement from all countries and monitoring their progress and achievement has become a priority. However, achieving these goals is an overall commitment requiring contributions from the various actors at supra-national, regional, national and local levels. In addition to various levels of action, there is also a wide distribution of actor types: international or regional organizations, national governments, education/ research institutions and NGOs. Thus, SDGs call for a global partnership at all levels, between countries and actors to work together and to achieve them (Caiado et al., 2018).

Given the multitude and variety of actors, it is of great importance to be able to make sense of what is being done towards the achievement of these goals and how these different actions fit together. For instance, since 2016, the SDG Index is being elaborated with the aim of evaluating the achievement of each goal and obtaining information from countries. This allows for the identification of priorities for action, supporting discussions/debates and identifying gaps in the data. However, the different indicators in this tool lack the ability to estimate to what extent other inputs (e.g. scientific output or research projects) are aligned with the SDGs. To achieve this, it is first important to recognize SDG-relevant content among vast quantities of text data, such as policy documents originating from various institutions or research content originating from academic institutions. This step is crucial, as it is a necessary per-condition to carry out any deeper analysis of what has been done towards the achievement of the SDGs.

To date different attempts in previous initiatives have been conducted. However, to the best of our knowledge, a single truly general purpose tool capable to classify text data from various sources and of various types (research output, patents, policy papers, etc.) has been lacking. In this paper, we present our answer to this problem – the OSDG platform and tool, which can identify SDG-relevant text and assign it to a specific SDG.

This paper is structured as follows: the first part provides a brief overview of existing research and other attempts to classify text to SDGs together with the main shortcomings of current approaches. In the second part, we present the idea behind the OSDG tool and how it addresses the shortcomings of the previous approaches. In the third part, we present the OSDG methodology and in the fourth and the fifth part - provide a brief description on how the tool can be used and the contribution policy. We conclude with further plans for project development and the next steps.

# 1. Overview of Existing Research Projects and Initiatives

The majority of the previous efforts to classify textual data to the 17 SDGs fall into the four main approaches:
- Ontology building;
- Supervised Machine-learning;
- Unsupervised machine-learning;
- A combination of the above approaches.

**Ontology building.** As an example of ontology building effort, Bautista-Puig and Mauleon (2019) created a [dataset](#) from a corpus of documents which had attached SDG labels and built an ontology of keywords for each SDG based on the the UN-descriptions and keywords extracted from the dataset. They also checked interrelations of topics and goals based on a co-occurrence map. Meanwhile, Buttigieg, McGlade, and Coppens (2015) have built an [ontology](#) from the legal documents connected to SDGs adopted by the UN. Another example of ontology-based classification is "LinkedSDG" project from the Statistics Division and the Division for Sustainable Development Goals' (DSDG) of the Department of Economic and Social Affairs, which extracts concepts from the texts and attempts to classify them based on the concept distribution (LinkedSDG 2020).

All ontology building exercises have several common features – all rely on either manually extracting features from unstructured data or curating the set of features extracted by a semi-automated approach (e.g. term co-occurrence matrix or TF-IDF vectors). Their main output is a (hierarchical) set of key-terms linked to a higher-level concept, in this case an SDG.

**Supervised machine learning.** Supervised machine learning approaches work by taking a set of input data together with the output labels and using a dedicated algorithm to find the best way to match the input data with the outputs, so that such trained models can be used to make predictions on new unseen data. For instance, Pincet, Okabe and Pawelczyk (2019) took a set of OECD documents with ex ante known SDG labels and developed a predictive model using Random Forests and XGBoost algorithms. This model was then used to analyse and map the documents related to Official Development Assistance (ODA) to SDGs. As another example, a deep learning model utilising term and document embeddings was used to detect UN General Assembly (UNGA) resolutions implicitly related to SDGs (Sovrano, Palmirani & Vitali 2020). The model was trained on another set of UNGA resolutions mentioning the SDGs explicitly. Another example is Wastl et al., (2020), that trained a machine-learning model using a training dataset from Digital Science "Dimensions" database. However, this is a proprietary tool not available to the general public.

The main aim of supervised machine learning approaches is to develop a model which would perform well on new and unseen data and it is the model (not the feature sets) that is the main output of these efforts. The main problem with these approaches is that the available labelled data for SDGs often comes from highly specific homogeneous corpora, such as UNGA resolutions or OECD documents. This always poses a danger that a model will pick up a number of corpus-specific features, which negatively affects model performance on out of corpus data (Van Asch & Daelemans 2010). Additionally, while it is possible to relatively easily merge, combine and extend different ontologies, the same cannot be said for the supervised ML models - the only way to change or adjust the model is to either train it on an adjusted dataset or re-train using tweaked model hyper-parameters.

**Unsupervised machine learning.** This represents kind of a middle ground between the two previous approaches. On one hand, it relies on dedicated algorithms to identify the most salient features in data. However, in this case the aim is not to predict a specific label, but rather find the underlying structure/clusters in the data. The features from these clusters can then be (manually) mapped to labels, de facto producing something very similar to ontologies. An example of such effort is the SDG Pathfinder project developed by the OECD and pooling together documents from several international Organizations (OECD 2020). Topic modelling is applied to the pooled corpus of documents and then the topics are matched to different SDGs. The result is both: a meaningful list of topics linked to SDGs as well a model which allows to detect these topics in new text and thus link new documents to SDGs.

**Hybrid approaches.** Finally, the hybrid approaches combine elements from several of the above approaches into a comprehensive methodological framework. An example of such endeavour is the work carried out by SIRIS group (Duran-Silva, Fuster, Massucci & Quinquillà 2019) in "[Science4SDGs](Science4SDGs)" project. In their approach, an initial set of ontology items is enriched using a "WORD2VEC" model by finding similar items in a larger corpus of relevant documents and by cross-referencing the initial "seed" ontology with other existing ontologies. Another example is Nakamura et al. (2019) which identified a set of SDG papers (the core of SDG documents and expanded based on citations) and applied a clustering algorithm based on the bibliographic coupling to create a SDG-topic map. This project relies on a proprietary "Web of Science" database and the associated "Dimensions" analytics tool. Another advanced project applying such methodology is undertaken by the Aurora Network which aims to classify the scientific output of HEIs into the different SDGs (Vanderfeesten & Otten 2017). Their effort is based on expert-crafting keyword queries for each of the 17 SDGs. Scientific publications are fetched from bibliometric and document databases using these queries and assembled into a labelled corpus. Moreover, their results are validated through an expert pool. In the next stage of the project, the plan is to develop a supervised ML model using said corpus as a training set.

There are many research initiatives that seek to contribute to the general goal of developing a robust and comprehensive approach to classify text data to Sustainable Development Goals. However, many of them have certain weaknesses and shortcomings. Ontologies, though high-quality are often not very comprehensive and cannot claim to capture the totality of the SDG related discourse. Supervised machine learning models are usually trained on relatively small and homogeneous corpora and, consequently struggle with out-of-sample cases. Furthermore, models trained on different sets of data with different parameters are almost impossible to integrate. Unsupervised machine learning holds a great promise, but so far has been applied to a relatively small corpora of text (e.g. SDG Pathfinder).

Seeing this urged us to develop the OSDG (beta) platform. OSDG platform provides a way to integrate several existing approaches into a single framework as well as giving the means to develop it further by a dedicated research community.

## 2. OSDG Project

The OSDG presents an opportunity to integrate data from multiple sources into a single framework for SDG classification. Our aim is to build a comprehensive framework that re-uses and integrates existing knowledge, as opposed to building yet another ontology from scratch.

The OSDG was specifically designed with this goal in mind, i.e. it has the ability to integrate data from multiple data sources and obtained using different methods. This project is fully open, which means that everyone can access the underlying data and scripts that make the tool work. Researchers are welcome to contribute their own data and suggestions for improvements to make this tool better.

Currently, the OSDG is maintained by the [Public Policy and Management institute (PPMI)](#) and [Technote.ai](#) with the plans to hand it over to an independent body if it builds momentum. Section 5 of this paper details out some potential steps in this direction.

This project provides important orientation in terms of Research, Development and Innovation (R&D+i) to respond to major societal challenges and could be useful for the policymakers or any stakeholder to quantify the efforts on this topic.

# 3. Methodology

OSDG integrates the existing research into a comprehensive approach and does so in a way that avoids the weaknesses of the individual approaches and the duplication of the research efforts. In short, the OSDG builds an integrated ontology from the feature sets identified in previous research and then matches the ontology items to the topics from Microsoft Academic (Sinha et al. 2015 ; Wang et al 2019). See Figure 1.

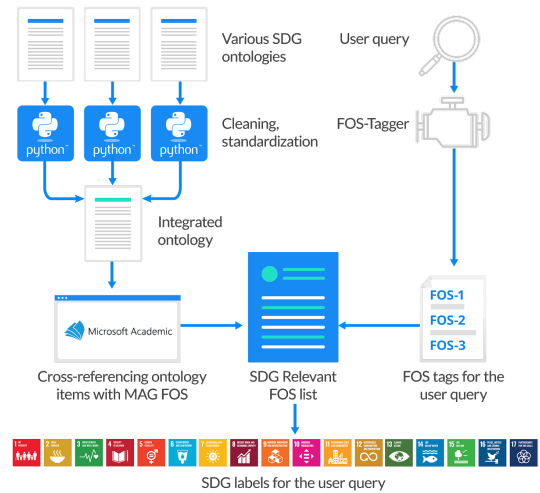

**Figure 1: OSDG Methodological Schema ([higher resolution here](#))**

In other words, we take relevant text features from the previous research, clean them and then merge them into a comprehensive OSDG ontology. From ontology-building approaches we took individual ontology items, from the supervised machine learning models we took the most significant/important features for each SDG label (e.g. beta coefficients from the logistic regression models). From unsupervised machine learning efforts we took the most important words for each topic.

This allowed us to create a comprehensive SDG ontology (containing >14K key-terms at the time of writing and constantly growing). However, in order to be useful, this ontology needed to be operationalised. To do this, we mapped the OSDG ontology items to the Fields of Study in Microsoft Academic Graph (MAG) (Sinha et al. 2015 ; Wang et al 2019) (see [here](#)). Microsoft research has performed hierarchical topic modeling over a gigantic corpus of scientific publications (more than 235 M at the time of writing). This produced more than 700 K hierarchically-linked topics (see here). Our mapping was done by linking the OSDG ontology items with similar FOS names. We considered an ontology item and FOS name to be similar if their Levenstein similarity ratio exceeded 0.85. By doing this, we achieve two things:

- Expand the ontology - acquire more key terms associated with the relevant MAG Topics or as they a nativelly called - Fields of Study (FOS);
- Capture more nuanced relationships between individual terms. For instance, the term "oil" is very relevant for a number of SDGs. However, it is truly relevant for SDGs when it is being mentioned in the context of fossil fuel reduction or "post-oil energy" (Spellman 2016). Naturally, it "oil" is mentioned in the context of novel drilling techniques it is not SDG-relevant. Topics from topic modelling allow us to capture these different contexts bringing this nuance to text classification.

The final component of the methodology is the tool that assigns Fields of Study to a piece of text. Since we do not have access to the original MAG topic model, we calculated TF-IDF vectors for each of the MAG FOS and developed a tool that assigns FOS based on the cosine similarity between the input vector and FOS vectors. This approach assigns very similar FOS tags to texts, though, naturally, these are not fully identical to the original MAG FOS Tags.

When a user posts a query to the OSDG tool, it does the following:

1. (optional) If a user post a DOI number, the tool looks up the text of the scientific abstract based on the DOI number;
2. Text (user input or scientific paper abstract retrieved using DOI) is tagged with FOS'es using the FOS-tagger;
3. FOS'es assigned to the text are cross-referenced with the OSDG Ontology; FOS overlap between the text and each SDG is calculated and returned to the user;
4. In the OSDG web tool, the numeric value of the relationship between the SDG and text is interpreted as either "Strong" or "Moderate". This depends on the SDG-specific threshold. The threshold value can be found on the GitHub repository.

We maintain an up-to-date list of data sources that are used to build the ontology as well as copies of the input datasets, processing scripts and the resulting combined ontology in the project GitHub repository. The datasets used in the OSDG version 0.0.2 can be found in Table 1.

**Table 1: Data sources used in OSDG v0.0.2 (Most recent version here)**

|    | Dataset name | Source |
|----|---|---|
| 1. | SDG Ontology compiled by Dr Nuria B. Puig and E. Mauleon | Link |
| 2. | Mapping from "FP7-4-SD" Project (edited by PPMI) | Link |
| 3. | "SDG Pathfinder" Project | Link |

| 4. | "Linked SDGs" Project | [Link](#) |
|---|---|---|
| 5. | "SDGIO" Project | [Link](#) |
| 6. | Keywords from the documents of the European Commission | [PPMI](#) (*Skrynik & Stanciauskas 2020 forthcoming*) |

# 4. Using and Contributing to OSDG

We offer two ways how researchers can use OSDG tool. First, users can use the OSDG website ([https://technote.ai/osdg](https://technote.ai/osdg)), which allows users to SDG-tag documents via short text (e.g. an abstract of a scientific paper or summary of a policy document) or via DOI number. The website allows tagging documents by the DOI number individually or in bulk.

We also offer more tech-savvy users to download and run dockerised version of the OSDG API, which can be found on the repositories in [GitHub](#) and [DockerHub](#). The dockerised version of the app supports short text tagging only.

The project repository on [GitHub](#) also houses the complete project dataset:
1. Input datasets;
2. Data cleaning and preprocessing scripts;
3. Complete combined OSDG Ontology;
4. Linking to MAG.

**We welcome user contributions on all aspects of the OSDG platform.** The easiest way to contribute to OSDG is to use the [web platform](#). There after SDG-tagging individual texts users are offered to suggest more appropriate SDG tags, if they are unhappy with the output of the OSDG tool. These suggestions will be pooled together and integrated into the OSDG as a separate data source (available on [GitHub](#)), so the tool can improve with use. **We highly encourage all users to provide feedback to the tool as this data is crucial for further improving the SDG labels.**

Contributions to the items listed above can be made by creating a pull request on [GitHub](#). By doing so, researchers can suggest new data sources or improve the current procedures for cleaning/processing data or building the ontology.

To contribute to the FOS-tagging engine, please contact the OSDG project maintainers through the OSDG website ([https://technote.ai/osdg](https://technote.ai/osdg)).

# 5. Next steps in OSDG

The vision is that the OSDG will become an ever growing ontology and framework for SDG classification. It will be used to classify publications, but also technological outputs (patents), research projects and even the outputs organisations, such as the Organisation for Economic Co-operation and Development (OECD), the European Commission (EC), United Nations (UN), etc. The framework will span across sectors, scientific domains and countries/languages.

We expect the OSDG to receive both interest and criticism. The latter is needed to identify and solve weaknesses in the overall approach or detailed predictions that the tool makes. The 'beta' label implies that the OSDG has a way to go before it is mature.

We invite interested parties to collaborate. Following the launch of the beta version there will be a 6-month period (until December 31st 2020) to express interest and join a taskforce for OSDG development.[3] Then, approximately by the end of 2020, an OSDG White Paper will be published outlining key priorities for the next few years. An OSDG Board will be formed to oversee the implementation of the priorities.

Some potential areas currently considered by PPMI and Technote researchers include:
- Integration of more data sources beyond the currently used ones, especially those beyond publications and policy documents
- Reconsider the overall threshold to assign SDG relevance; find a more appropriate balance between precision and recall;
- Systemic cross-validation of the data produced using two or more different approaches by independent teams
- Completing the linking of MAG Fields of Study to the SDGs
- Linking of SDGs to other frameworks, e.g. the Frascati Manual, EU and national policy objectives, research missions, etc.

You are invited to share your interest, ideas or feedback to the OSDG project on [GitHub](#) or via email ([osdg@technote.ai](mailto:osdg@technote.ai)). The OSDG is open to researchers from various domains, particularly those interested in the topics of sustainable development, scientometrics, natural language processing, machine-learning, Big Data or AI.

---

[3] To do so, write to [osdg@technote.ai](mailto:osdg@technote.ai)

## Links

- OSDG Platform : https://technote.ai/osdg ;
- OSDG on Github : https://github.com/Technote-ai/osdg ;
- OSDG on DockerHub : https://hub.docker.com/r/technoteai/osdg